\documentclass[runningheads]{llncs}

\usepackage[letterpaper]{geometry}
\usepackage{tikz}
\usepackage{amsmath}
\usepackage{graphicx}
\PassOptionsToPackage{hyphens}{url}\usepackage[colorlinks=true, allcolors=blue]{hyperref}
\usepackage[strings]{underscore} 
\usepackage{enumitem}
\usepackage{subcaption} 

\def\commentson{1} 

\ifnum\commentson=1

\newcommand{\mahdi}[1]{{
		{\color{red}\textbf{Mahdi:}
			#1
		}\normalcolor}}

\newcommand{\catherine}[1]{{
		{\color{magenta}\textbf{Catherine:}
			#1
		}\normalcolor}}

\newcommand{\panos}[1]{{
		{\color{blue}\textbf{Panos:}
			#1
		}\normalcolor}}

  \newcommand{\srini}[1]{{
		{\color{blue}\textbf{Srini:}
			#1
		}\normalcolor}}

\else

\newcommand{\mahdi}[1]{}

\newcommand{\catherine}[1]{}

\newcommand{\panos}[1]{}

\newcommand{\srini}[1]{}

\fi

\title{Privacy-Enhancing Technologies for\\ Financial Data Sharing}
\author{Panagiotis Chatzigiannis, Wanyun Catherine Gu, Srinivasan Raghuraman, \\Peter Rindal, Mahdi Zamani}
\institute{Visa Research, Palo Alto, CA}
\date{March 3, 2023}

\begin{document}
\sloppy
\maketitle

\begin{abstract}
Today, financial institutions (FIs) store and share consumers' financial data for various reasons such as offering loans, processing payments, and protecting against fraud and financial crime.
Such sharing of sensitive data have been subject to  data breaches in the past decade. 

While some regulations (e.g., GDPR, FCRA, and CCPA) help to prevent institutions from freely sharing clients' sensitive information, some regulations (e.g., BSA 1970) require FIs to share certain financial data with government agencies to combat financial crime. 
This creates an inherent tension between the privacy and the integrity of financial transactions. In the past decade, significant progress has been made in building efficient privacy-enhancing technologies that allow computer systems and networks to validate encrypted data automatically.

In this paper, we investigate some of these technologies to identify the benefits and limitations of each, in particular, for use in data sharing among FIs.
As a case study, we look into the emerging area of Central Bank Digital Currencies (CBDCs) and how privacy-enhancing technologies can be integrated into the CBDC architecture. Our study, however, is not limited to CBDCs and can be applied to other financial scenarios with tokenized bank deposits such as cross-border payments, real-time settlements, and card payments.
\end{abstract}

\section{Introduction}

In the modern economy,  services for all types of financial monetary transactions are provided by a number of FIs. Naturally, those FIs need to handle a huge amount of data, especially as the economy gradually shifts from cash towards transactions in electronic format. At the same time, FIs typically operate in a heavily-regulated environment in order to incorporate laws related to consumer protection, data privacy, and financial crime.

With the sheer amount of data available to FIs, which includes personally-identifiable information, there is a need to utilize this data towards providing better financial services, enforcing regulations and laws, and identifying financial transactions associated with illicit activities, e.g., money laundering, tax evasion, and terrorism financing~\cite{Financia4:online}. This requires a collective effort on behalf of FIs in order to better achieve those goals. However, FIs cannot simply share the data with each other, as this would be against regulations on privacy protection in many jurisdictions~\cite{ccpa,cdpa}. In addition, such laws that have a much broader reach~\cite{GDPR} make nearly all forms of private data sharing even more restrictive. 

While there are many ways of sharing data in a privacy-preserving way (e.g., utilizing techniques such as encryption or differential privacy), it is challenging to design algorithms that perform  computation over those data and still be reliable and efficient enough to preserve data utility. In other words, the algorithms that work on data processed to preserve privacy should closely mimic the properties of algorithms that would work on unprocessed data. For example, suppose FIs were sharing private data among themselves with no restrictions, and one of them computed a customer's credit score. When enforcing privacy-preserving financial data sharing, the algorithm collectively employed by FIs should ideally output the same credit score within the same time period.


In this paper, we focus on the problem of privacy-preserving data sharing between FIs, and not between FIs and individual clients (where for example, a client would be required to prove to an FI the source of his or her fund). While these cases are commonly encountered on a client level for a bank, these could be eventually replaced by data sharing between FIs in a recursive fashion, similar to established standards designed to prevent tax evasion \cite{crs}.

We identify two main classes of such algorithms that can perform these functionalities: multi-party computation (MPC) and federated learning (FL). To showcase their capabilities and limitations, we consider central bank digital currencies (CBDCs) as a payment system model. However, this consideration does not imply a loss of generality and our discussion applies to mainstream financial systems as well, such as international bank transfers, ACH transactions, and credit cards. In addition, we briefly discuss applications of zero-knowledge proofs (ZKPs) and homomorphic encryption (HE) in the general paradigm of privacy-preserving data sharing that improve the security and efficiency of MPC and FL in certain use cases.

\paragraph{Related Work.} Two whitepapers published by the World Economic Forum and Deloitte in 2019~\cite{deloitte,wef} provide a high-level overview
of privacy-enhancing technologies (PETs) and their potential value in a financial institution setting. In our work we provide a more rigorous and in-depth  analysis of how those PETs can be utilized specifically for cross-FI data sharing, while differentiating between use-case goals where one technology might be more suitable over another, considering the recent advances in the fields of cryptography and machine learning. Finally, we also propose a new paradigm, which we refer to as PET-as-a-service (PETaaS) paradigm, to address the challenges for businesses and FIs to perform data sharing with currently available PETs.

%
%

\subsection{How Much Privacy Is Needed?}
\label{subsect:neededprivacy}
In a financial system, a client traditionally establishes a ``custodial" relationship with an FI (e.g., a bank or a broker) which creates an account associated with some credentials for the client to manage their assets, initiate payments or transfers from and to their accounts, etc.
A digital form of credentials allows the client to prove their identity when accessing the system and usually comes in the form of username-password pairs, biometrics, credit or debit cards, or digital tokens stored on smartphones. 

While credentials serve as the client's primary means to access their account, the custodian can generally access the data directly without using the credentials. Also, the custodian often shares certain account data with third parties to use their services, e.g., to process payments and loans, obtain credit reports, or audit transactions. However, the custodians generally do not share the credentials or the ability to access the accounts with third parties in an unrestricted way.\footnote{Third-party access to user account data is usually provided in a limited form with restrictions set by client-approved policies.}

	

Depending on the level of trust in custodians, a client may be offered two levels of privacy. In some cases, an \emph{end-to-end privacy}\footnote{Sometimes referred to as \emph{full} or \emph{user-level} privacy.} could be offered, where no intermediary (including the custodian) is allowed to access the transaction data, except the transacting parties. This is the strongest level of privacy and is increasingly becoming desirable. Notably, several payment systems in the context of cryptocurrencies already offer such strong level of privacy \cite{SP:BCGGMT14,FC:BAZB20,van2013cryptonote} while there is significant research effort towards achieving regulatory compliance of such systems \cite{a16zzk,ACNS:ChaBalCha21}.

Offering privacy at the user level, however, often dramatically impacts the performance of the financial system which could jeopardize its usability and cost-effectiveness altogether. This is primarily because financial custodians are often heavily involved in every transaction from authentication to authorization and settlement, and performing such functionalities over hidden data on a frequent basis entails extensive use of advanced cryptographic tools that make computation over hidden data possible.
While occasional usage of such methods might be viable from a practical standpoint, extensive use of them typically imposes severely higher communication bandwidth, latency, and computational costs.\footnote{While advanced cryptographic methods have become significantly more efficient, especially in the past decade, they are still, unfortunately, far from being used in real-time to guarantee end-to-end privacy.}

In this paper, we focus on a weaker level of privacy, which we refer to as the \emph{cross-institutional privacy}. In this approach, the custodian may store its client's data in an unencrypted, but access-controlled data store, while occasionally sharing account data with third parties in an encrypted format. To achieve this, the custodian uses advanced cryptographic methods 
to share sensitive data in a manner that prevents any leakage of information to the other institution.

While this still does not account for real-time sharing of data across institutions in a privacy-preserving fashion due to the large overhead of such cryptographic methods, we observe that most data-sharing use cases in the financial world, as we explore in the next section, require only occasional sharing of sensitive data with other institutions.

Finally, privacy can be distilled into \emph{identity privacy} (or \emph{anonymity}) and \emph{transaction privacy} (or \emph{confidentiality})~\cite{ACNS:ChaBalCha21,allen2020design}. The former is about hiding any direct information related to the  identity of transacting users, e.g., name and address. In addition, data that can be used or intersected with other data and infer a user's identity, e.g., an account number or credit card number, are usually referred to as \emph{personally-identifiable information~(PII)}, and under identity, privacy should be hidden as well.
The latter is about hiding details on the financial transaction, e.g., amounts, currency, and location. Depending on the use case, there could be different levels of sensitivity for different data attributes leading to a choice to guarantee either identity and transaction privacy or both of them.

\subsection{Data Sharing Use Cases in the Financial World}  
We now discuss a few use cases, where FIs and governments could benefit significantly from sharing data across institutions to obtain useful insights into financial transactions. Such sharing practices have traditionally been either avoided due to restrictions enforced by privacy regulations, competitive disadvantages, or major data breaches due to improper handling of data and privacy measures.
The common goal across the use cases reviewed in this section is to incentivize FIs to improve their risk management mechanisms, and therefore to reduce costs, by participating in safe data sharing practices 
with other institutions without compromising regulatory and competitive measures and with minimal operational overhead.

\subsubsection{Fraud Detection.} Financial fraud happens when an individual, usually other than the legitimate owner of an account, initiates a transaction from the account to purchase goods/services without the owner's permission, or to steal assets from the owner or the financial intermediaries.\footnote{Examples of financial fraud include credit/debit card fraud, account takeover, violation of chargeback or refund policies, etc.} FIs usually detect fraud based on anomalies observed in contextual information collected from the owner and past transactions, e.g., the location, time, frequency, vendors involved, etc. Typical techniques to detect fraud include building decision trees and machine learning models that are trained based on past cases of fraud. While abnormality testing usually happens in real-time at the time of the transaction, aiming to detect misuse of the financial system such as credentials abuse, abnormal access patterns, etc.,
training of fraud models happens mostly offline.\footnote{The frequency of incorporating a transaction into the fraud model can range from days to months after the transaction.} Most FIs today rely on their own data collected from past transactions. However, complex fraud models could be created by sharing data and/or models across institutions to improve accuracy. 

\subsubsection{Financial Crime Detection.} Usually passive checks are done over many transactions to detect systematic abuse of the financial systems such as money laundering and terrorism financing. As a reference, \cite{DBLP:conf/IEEEares/KolachalaSAV21} 
suggested some measures and challenges for implementing anti-money laundering mechanisms in the context of cryptocurrencies. 

\subsubsection{Credit and Activity Monitoring.} Some FIs offer credit reports to other FIs for purposes such as loans and mortgages~\cite{hurley2016credit}. Such reports are often created based on the information shared by banks about the consumer's financial diligence in paying bills, etc. Also, some FI may provide consumer activity information such as payment time and location to other institutions for  advertisement or recommendation services. 

\subsubsection{Macroeconomic Statistics.} Economic indicators such as the consumer price index (CPI) and gross domestic product (GDP) are used to inform policymakers in setting policy rates and macroeconomic agendas. These indicators are imperfect since the data collection process is largely manual~\cite{cpi} and may involve collecting sensitive information from consumers and businesses.\footnote{For instance today, there is typically a two-week lag to publish the monthly CPI to measure the rate of inflation, and there is a one-month lag to publish the first estimate of the quarterly GDP.} Aggregating data collected from multiple institutions on a frequent, automatic basis would result in significantly richer macroeconomic indicators. However, such frequent sampling of data from various sources could create consumer and cross-institutional privacy challenges. Nevertheless, private sampling of data is susceptible to malicious bias in input data that could affect the accuracy of aggregation. Detecting and eradicating such biases is challenging.   

	


\subsection{CBDC Model and Features}
CBDCs are digital forms of central bank money (i.e., money that is a liability of the central bank) \cite{cbdc-fedreserve}. 
The motivations driving central banks to explore CBDC are varied. For CBDC to be adopted as a medium of exchange in various economic activities, one use case with substantial potential is to have CBDC as a macroeconomic tool in monitoring and aggregating macroeconomic activities in close to real-time, given the entirely digital nature of CBDC. Economic indicators today that are used to inform policymakers in setting policy rates and macroeconomic agenda are imperfect since the data collection process at initiation tends to be largely manual~\cite{cpi} and raw data that is being collected is not private and may contain sensitive information.

\begin{figure}
	\centering
	
	\includegraphics[width=0.8\linewidth]{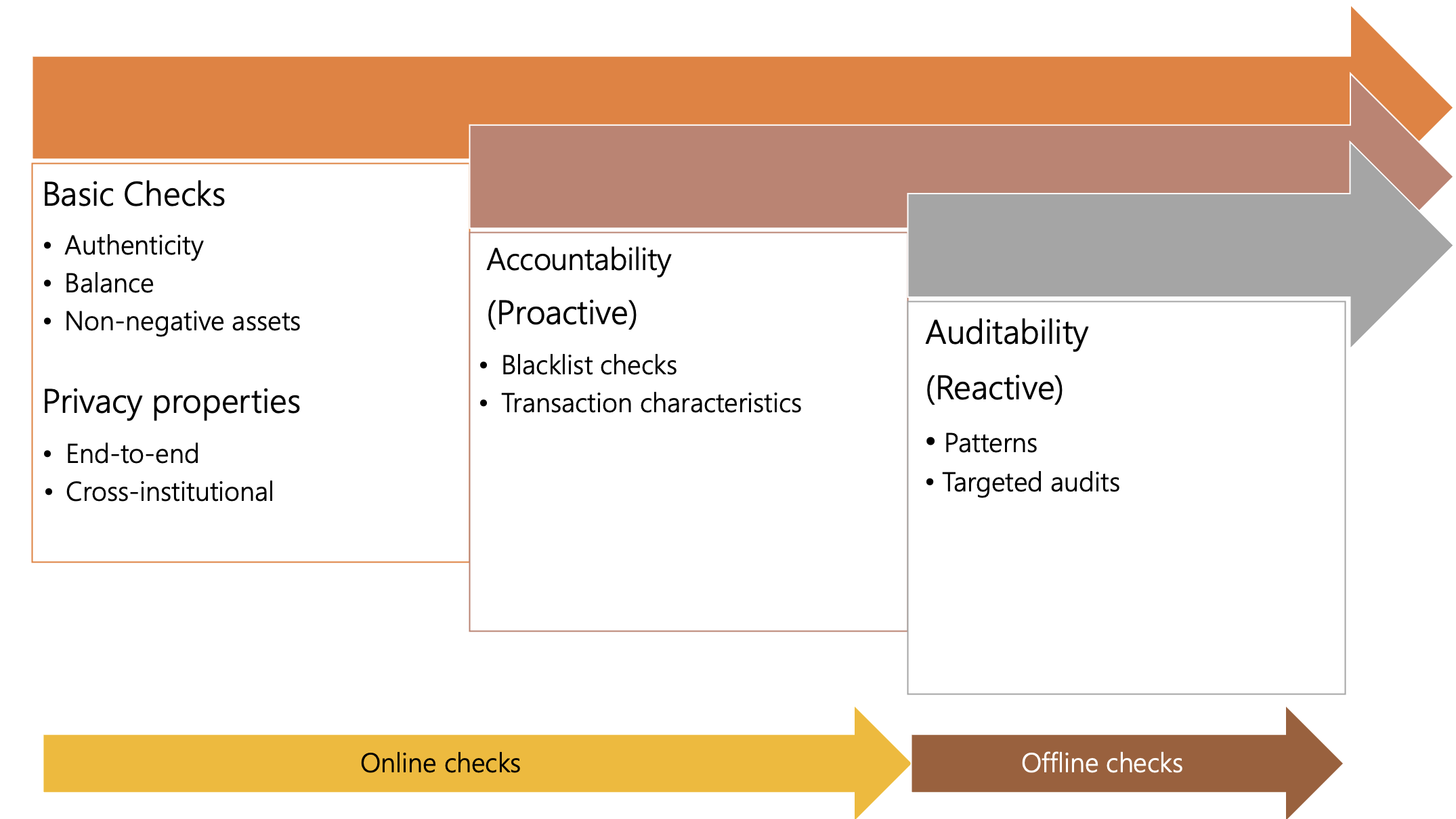}  \caption{Payment Verification Flow}
	\label{fig:real-offline}
\end{figure}

Fundamentally, a CBDC system needs to satisfy basic security properties to ensure authenticity, uniqueness, and theft prevention of the tokens it circulates. In addition, the system might consider some level of privacy for its users to protect against individual citizen tracking  and human rights violations.
However, the CBDC system must also be able to facilitate compliance checks, which will prevent malicious or illicit activities such as fraud or money laundering. These checks can be proactive and executed at the time of the transaction (e.g., checking against blacklists or deciding on transaction characteristics that might indicate fraud), or reactive and executed in a deferred approach, because of a more computationally intensive process (e.g., detecting patterns at a transaction level that might indicate money laundering), as shown in Figure \ref{fig:real-offline}. Naturally, this creates a tension between user privacy and lawful auditability, which has been thoroughly discussed and analyzed both in a decentralized cryptocurrency~\cite{ACNS:ChaBalCha21} and a CBDC setting~\cite{allen2020design}. 
Another desirable feature of a CBDC system that often creates tension with privacy is to employ account metadata tracking functionalities for providing auxiliary services to customers (e.g., creditworthiness or personalized ads).


	

	

The above goals and features, such as Banks learning which of their customers are fraudsters, which transactions indicate money laundering activities, or what is the credit score of each client, require the commercial Banks participating in the CBDC system to collectively work and share their data in a privacy-preserving way, i.e., without a commercial Bank exposing its data to other commercial Banks.
Also note that the scope of financial data sharing is not limited to a single CBDC system, but can extend also to cross-border payments. However, this scope excludes the central Banks themselves, as they are oblivious to client account or transaction information, and are solely responsible for CBDC issuance and other high-level monetary policies.

\section{PETs for Financial Data Sharing}

We now briefly describe a set of advanced cryptographic methods that can be used to construct privacy-enhancing technologies for financial data sharing. Such technologies often rely on one or more computer nodes working together using a \emph{protocol} to hide data and execute a functionality over the hidden data.
We will later discuss how one may combine some of these methods to create a hybrid solution.


\subsubsection{Adversarial Model.} Protocols are often characterized by their ability to resist cyberattacks against the protocol's participating nodes. If we consider a participant (aka, an \emph{adversary}) who might attempt to breach the privacy of the system but without deviating from the protocol steps, then we refer to it as an \emph{honest-but-curious} (aka, \emph{semi-honest}) adversary. However, if we consider an adversary (participant or third-party) who might actively attempt to corrupt the protocol, for example, by sending specially crafted messages in order to prevent the protocol from reaching its goals, then we refer to it as a \emph{malicious} (aka, \emph{Byzantine}) adversary.

\subsection{Secure Multi-Party Computation}
\subsubsection{Background.} 
Secure multi-party computation (MPC)~\cite{STOC:CFGN96,FOCS:Yao82b} is a family of cryptographic methods which enables parties to jointly compute a function, where each party provides a private input. The requirement is that, among others, the input is not revealed to anyone, but also no information related to that input is leaked due to the protocol execution, apart from what can be inferred from the desired output.
As an example, consider a set of $n$ employees who want to compute the average salary, but without revealing their individual salaries to other parties. In this case, the private inputs to the MPC will be the $n$ individual salaries, and the output is the average salary of all employees.

Typically, MPC uses a \emph{secret sharing} method, which privately distributes a secret input among a group of computing nodes. In a secret sharing scheme such as that of Shamir \cite{DBLP:journals/compsec/DawsonD94}, the secret is mathematically split into pieces each of which does not reveal any information about the input but an authorized subset of shares could be used to reconstruct the input. In the employee example above, each employee would split the private individual salary number into $n$ pieces, with all pieces adding up to the salary, and distribute $n-1$ pieces to the other parties. The average can then be computing using all of the pieces.

\subsubsection{MPC in Financial Data Sharing.}
The first general approach to enable financial data sharing in a privacy-preserving way is to have banks engage in an MPC protocol. The private inputs into the protocol will be each bank's private ledger, and the outputs will depend on the initial goal (e.g., if a pattern was detected that is an indication of a fraudulent transaction and/or a malicious client or a number that indicates the client's creditworthiness). As an alternative approach, the banks can outsource the MPC computation to the CBDC DLT validator committee. In this outsourcing mechanism, the banks can secret-share their private data with the validator set, and the privacy of the data will be preserved as long as at least one validator behaves honestly. Note that the second approach is preferable when the number of participating banks is much larger than the number of validators, as the efficiency of the MPC decreases as the number of MPC parties increases. On the other hand, it would be preferable to have the banks directly participate in the MPC protocol if the individual bank's data is very large, which would make the secret-sharing process expensive, unless the banks can pre-process their data into a smaller aggregate (a survey of MPC techniques when the number of input providers is large can be found in \cite{EPRINT:SaiZam14}). Other examples include the recent developments include highly flexible frameworks for performing database joins and aggregations on encrypted data \cite{DBLP:conf/ccs/BadrinarayananD22}, capable of joining/processing fully encrypted datasets with millions of records in a few seconds. Related techniques such as private set intersection are even faster, computing the common identifiers between datasets with two million records in less than a second \cite{blazing}, without revealing the datasets themselves.

As an example, an MPC protocol that could be used towards sharing financial data in a privacy-preserving way is GraphSC \cite{SP:NWIWTS15}, a secure parallel programming paradigm that enables secure computation on large datasets, and particularly lends itself to efficiently implement oblivious versions of graph-based algorithms. GraphSC supports algorithms similar to ``gossiping" algorithms in the distributed graph algorithms literature, which proceed in several phases. In each phase, each node of a graph ``gossips'', i.e., sends a message to all of its neighbors. At the end of this phase, nodes aggregate the information they have received and compute the message they will send in the next round. GraphSC enables one to generically compile a non-private algorithm comprised of the above phases into a privacy-preserving protocol that implements the same algorithm. For instance, graph-based traversal algorithms such as breadth-first search and depth-first search can be implemented in this framework. 

We can now represent financial transactions as a graph structure, where the vertices represent the customers and the edges represent the transactions between them. 
At a high level, frameworks such as GraphSC can implement secure payment auditability tools which would allow multiple financial institutions to bring their transaction data (modeled as transaction graphs) and perform audits or regulatory checks on their combined transaction data (which can be modeled as a larger transaction graph, which is the union of the graphs held by each of the institutions), after using all information pertaining to the transactions, i.e., the edges in the graph, in a privacy-preserving way. Note that the combined graph is not learned by any participating party in the protocol, which is key for preserving privacy when sharing such financial data, which in this case is customer identity information and their transactions.

\begin{figure}
\begin{subfigure}[h]{0.23\linewidth}
\resizebox{\columnwidth}{!}{
\tikzset{every picture/.style={line width=0.75pt}} 

\begin{tikzpicture}[x=0.75pt,y=0.75pt,yscale=-1,xscale=1]

\draw  [fill={rgb, 255:red, 0; green, 0; blue, 0 }  ,fill opacity=1 ] (280.91,165.45) .. controls (280.91,162.94) and (282.94,160.91) .. (285.45,160.91) .. controls (287.96,160.91) and (290,162.94) .. (290,165.45) .. controls (290,167.96) and (287.96,170) .. (285.45,170) .. controls (282.94,170) and (280.91,167.96) .. (280.91,165.45) -- cycle ;
\draw    (285.45,165.45) -- (352.04,176.51) ;
\draw [shift={(355,177)}, rotate = 189.43] [fill={rgb, 255:red, 0; green, 0; blue, 0 }  ][line width=0.08]  [draw opacity=0] (8.04,-3.86) -- (0,0) -- (8.04,3.86) -- cycle    ;
\draw    (285.45,165.45) -- (345.87,225.39) ;
\draw [shift={(348,227.5)}, rotate = 224.77] [fill={rgb, 255:red, 0; green, 0; blue, 0 }  ][line width=0.08]  [draw opacity=0] (8.04,-3.86) -- (0,0) -- (8.04,3.86) -- cycle    ;
\draw  [fill={rgb, 255:red, 0; green, 0; blue, 0 }  ,fill opacity=1 ] (355,177) .. controls (355,174.49) and (357.04,172.45) .. (359.55,172.45) .. controls (362.06,172.45) and (364.09,174.49) .. (364.09,177) .. controls (364.09,179.51) and (362.06,181.55) .. (359.55,181.55) .. controls (357.04,181.55) and (355,179.51) .. (355,177) -- cycle ;
\draw  [fill={rgb, 255:red, 0; green, 0; blue, 0 }  ,fill opacity=1 ] (347.45,230) .. controls (347.45,227.49) and (349.49,225.45) .. (352,225.45) .. controls (354.51,225.45) and (356.55,227.49) .. (356.55,230) .. controls (356.55,232.51) and (354.51,234.55) .. (352,234.55) .. controls (349.49,234.55) and (347.45,232.51) .. (347.45,230) -- cycle ;
\draw  [fill={rgb, 255:red, 0; green, 0; blue, 0 }  ,fill opacity=1 ] (349.5,105.05) .. controls (349.5,102.54) and (351.54,100.5) .. (354.05,100.5) .. controls (356.56,100.5) and (358.59,102.54) .. (358.59,105.05) .. controls (358.59,107.56) and (356.56,109.59) .. (354.05,109.59) .. controls (351.54,109.59) and (349.5,107.56) .. (349.5,105.05) -- cycle ;
\draw  [fill={rgb, 255:red, 0; green, 0; blue, 0 }  ,fill opacity=1 ] (452.41,143.45) .. controls (452.41,140.94) and (454.44,138.91) .. (456.95,138.91) .. controls (459.46,138.91) and (461.5,140.94) .. (461.5,143.45) .. controls (461.5,145.96) and (459.46,148) .. (456.95,148) .. controls (454.44,148) and (452.41,145.96) .. (452.41,143.45) -- cycle ;
\draw    (352,230) -- (445.41,232.38) ;
\draw [shift={(448.41,232.45)}, rotate = 181.46] [fill={rgb, 255:red, 0; green, 0; blue, 0 }  ][line width=0.08]  [draw opacity=0] (8.04,-3.86) -- (0,0) -- (8.04,3.86) -- cycle    ;
\draw  [fill={rgb, 255:red, 0; green, 0; blue, 0 }  ,fill opacity=1 ] (448.41,232.45) .. controls (448.41,229.94) and (450.44,227.91) .. (452.95,227.91) .. controls (455.46,227.91) and (457.5,229.94) .. (457.5,232.45) .. controls (457.5,234.96) and (455.46,237) .. (452.95,237) .. controls (450.44,237) and (448.41,234.96) .. (448.41,232.45) -- cycle ;
\draw    (354.05,105.05) -- (450.18,139.98) ;
\draw [shift={(453,141)}, rotate = 199.97] [fill={rgb, 255:red, 0; green, 0; blue, 0 }  ][line width=0.08]  [draw opacity=0] (8.04,-3.86) -- (0,0) -- (8.04,3.86) -- cycle    ;
\draw    (456.95,143.45) .. controls (449.54,53.45) and (304.42,26.18) .. (285.73,158.9) ;
\draw [shift={(285.45,160.91)}, rotate = 277.39] [fill={rgb, 255:red, 0; green, 0; blue, 0 }  ][line width=0.08]  [draw opacity=0] (8.04,-3.86) -- (0,0) -- (8.04,3.86) -- cycle    ;
\draw    (285.45,165.45) -- (348.26,109.5) ;
\draw [shift={(350.5,107.5)}, rotate = 138.3] [fill={rgb, 255:red, 0; green, 0; blue, 0 }  ][line width=0.08]  [draw opacity=0] (8.04,-3.86) -- (0,0) -- (8.04,3.86) -- cycle    ;
\draw    (359.55,177) -- (446.41,227.49) ;
\draw [shift={(449,229)}, rotate = 210.17] [fill={rgb, 255:red, 0; green, 0; blue, 0 }  ][line width=0.08]  [draw opacity=0] (8.04,-3.86) -- (0,0) -- (8.04,3.86) -- cycle    ;
\draw    (452.95,232.45) .. controls (472.9,270.31) and (281.43,287.37) .. (285.38,171.75) ;
\draw [shift={(285.45,170)}, rotate = 92.9] [fill={rgb, 255:red, 0; green, 0; blue, 0 }  ][line width=0.08]  [draw opacity=0] (8.04,-3.86) -- (0,0) -- (8.04,3.86) -- cycle    ;

\end{tikzpicture}
}
\caption{Money laundering}
\label{fig:moneylaundering}
\end{subfigure}
\hfill
\begin{subfigure}[h]{0.31\columnwidth}
\resizebox{\columnwidth}{!}{
\tikzset{every picture/.style={line width=0.75pt}} 

\begin{tikzpicture}[x=0.75pt,y=0.75pt,yscale=-1,xscale=1]
	
	\draw  [fill={rgb, 255:red, 0; green, 0; blue, 0 }  ,fill opacity=1 ] (280.91,165.45) .. controls (280.91,162.94) and (282.94,160.91) .. (285.45,160.91) .. controls (287.96,160.91) and (290,162.94) .. (290,165.45) .. controls (290,167.96) and (287.96,170) .. (285.45,170) .. controls (282.94,170) and (280.91,167.96) .. (280.91,165.45) -- cycle ;
	\draw  [fill={rgb, 255:red, 0; green, 0; blue, 0 }  ,fill opacity=1 ] (189.91,77.95) .. controls (189.91,75.44) and (191.94,73.41) .. (194.45,73.41) .. controls (196.96,73.41) and (199,75.44) .. (199,77.95) .. controls (199,80.46) and (196.96,82.5) .. (194.45,82.5) .. controls (191.94,82.5) and (189.91,80.46) .. (189.91,77.95) -- cycle ;
	\draw  [fill={rgb, 255:red, 0; green, 0; blue, 0 }  ,fill opacity=1 ] (158.5,195) .. controls (158.5,192.49) and (160.54,190.45) .. (163.05,190.45) .. controls (165.56,190.45) and (167.59,192.49) .. (167.59,195) .. controls (167.59,197.51) and (165.56,199.55) .. (163.05,199.55) .. controls (160.54,199.55) and (158.5,197.51) .. (158.5,195) -- cycle ;
	\draw  [fill={rgb, 255:red, 0; green, 0; blue, 0 }  ,fill opacity=1 ] (254.95,289) .. controls (254.95,286.49) and (256.99,284.45) .. (259.5,284.45) .. controls (262.01,284.45) and (264.05,286.49) .. (264.05,289) .. controls (264.05,291.51) and (262.01,293.55) .. (259.5,293.55) .. controls (256.99,293.55) and (254.95,291.51) .. (254.95,289) -- cycle ;
	\draw  [fill={rgb, 255:red, 0; green, 0; blue, 0 }  ,fill opacity=1 ] (276.36,39.38) .. controls (276.36,36.87) and (278.4,34.83) .. (280.91,34.83) .. controls (283.42,34.83) and (285.45,36.87) .. (285.45,39.38) .. controls (285.45,41.89) and (283.42,43.92) .. (280.91,43.92) .. controls (278.4,43.92) and (276.36,41.89) .. (276.36,39.38) -- cycle ;
	\draw  [fill={rgb, 255:red, 0; green, 0; blue, 0 }  ,fill opacity=1 ] (406.36,151.88) .. controls (406.36,149.37) and (408.4,147.33) .. (410.91,147.33) .. controls (413.42,147.33) and (415.45,149.37) .. (415.45,151.88) .. controls (415.45,154.39) and (413.42,156.42) .. (410.91,156.42) .. controls (408.4,156.42) and (406.36,154.39) .. (406.36,151.88) -- cycle ;
	\draw  [fill={rgb, 255:red, 0; green, 0; blue, 0 }  ,fill opacity=1 ] (373.86,249.88) .. controls (373.86,247.37) and (375.9,245.33) .. (378.41,245.33) .. controls (380.92,245.33) and (382.95,247.37) .. (382.95,249.88) .. controls (382.95,252.39) and (380.92,254.42) .. (378.41,254.42) .. controls (375.9,254.42) and (373.86,252.39) .. (373.86,249.88) -- cycle ;
	\draw    (163.05,195) .. controls (202.45,165.45) and (227.25,154.33) .. (282.89,164.95) ;
	\draw [shift={(285.45,165.45)}, rotate = 191.28] [fill={rgb, 255:red, 0; green, 0; blue, 0 }  ][line width=0.08]  [draw opacity=0] (8.04,-3.86) -- (0,0) -- (8.04,3.86) -- cycle    ;
	\draw    (194.45,77.95) .. controls (236.36,66.67) and (274.83,109.19) .. (285.01,162.99) ;
	\draw [shift={(285.45,165.45)}, rotate = 260.24] [fill={rgb, 255:red, 0; green, 0; blue, 0 }  ][line width=0.08]  [draw opacity=0] (8.04,-3.86) -- (0,0) -- (8.04,3.86) -- cycle    ;
	\draw    (280.91,39.38) .. controls (310.55,53.29) and (319.24,124.81) .. (286.96,163.7) ;
	\draw [shift={(285.45,165.45)}, rotate = 311.52] [fill={rgb, 255:red, 0; green, 0; blue, 0 }  ][line width=0.08]  [draw opacity=0] (8.04,-3.86) -- (0,0) -- (8.04,3.86) -- cycle    ;
	\draw    (285.45,165.45) .. controls (324.65,136.05) and (369.66,134.55) .. (408.54,150.86) ;
	\draw [shift={(410.91,151.88)}, rotate = 203.8] [fill={rgb, 255:red, 0; green, 0; blue, 0 }  ][line width=0.08]  [draw opacity=0] (8.04,-3.86) -- (0,0) -- (8.04,3.86) -- cycle    ;
	\draw    (259.5,289) .. controls (232.9,254.03) and (229.11,188.5) .. (282.95,166.43) ;
	\draw [shift={(285.45,165.45)}, rotate = 159.56] [fill={rgb, 255:red, 0; green, 0; blue, 0 }  ][line width=0.08]  [draw opacity=0] (8.04,-3.86) -- (0,0) -- (8.04,3.86) -- cycle    ;
	\draw    (378.41,249.88) .. controls (324.05,245.05) and (289.69,235.2) .. (285.57,167.52) ;
	\draw [shift={(285.45,165.45)}, rotate = 87.08] [fill={rgb, 255:red, 0; green, 0; blue, 0 }  ][line width=0.08]  [draw opacity=0] (8.04,-3.86) -- (0,0) -- (8.04,3.86) -- cycle    ;
	\draw    (285.45,165.45) .. controls (234.54,185.1) and (208.07,191.73) .. (165.66,194.82) ;
	\draw [shift={(163.05,195)}, rotate = 356.1] [fill={rgb, 255:red, 0; green, 0; blue, 0 }  ][line width=0.08]  [draw opacity=0] (8.04,-3.86) -- (0,0) -- (8.04,3.86) -- cycle    ;
	\draw    (285.45,165.45) .. controls (235.29,149.9) and (208.84,117.19) .. (195.46,80.76) ;
	\draw [shift={(194.45,77.95)}, rotate = 70.84] [fill={rgb, 255:red, 0; green, 0; blue, 0 }  ][line width=0.08]  [draw opacity=0] (8.04,-3.86) -- (0,0) -- (8.04,3.86) -- cycle    ;

\end{tikzpicture}
}
\caption{Ponzi scheme}
\label{fig:ponzi}
\end{subfigure}%
\hfill
\begin{subfigure}[h]{0.39\linewidth}
	\resizebox{\columnwidth}{!}{
		\input{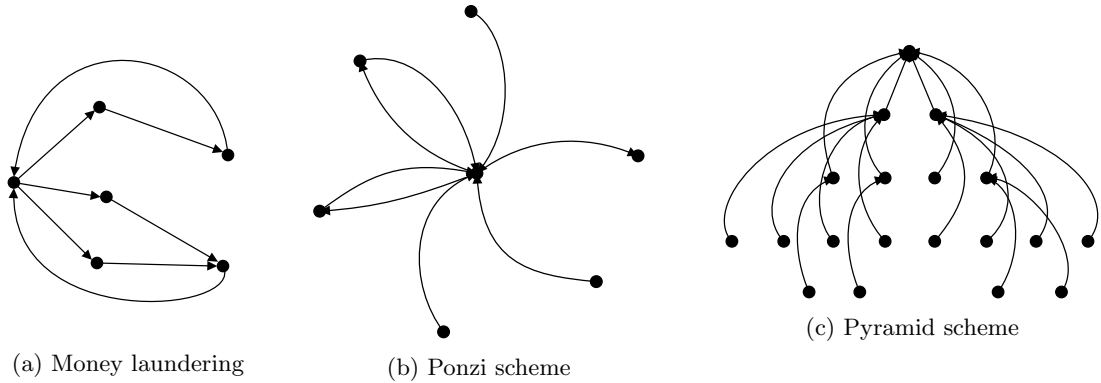}
	}
	\caption{Pyramid scheme}
	\label{fig:pyramid}
\end{subfigure}
\caption{Examples of transaction graph patterns related to potential fraud activity}
\label{fig:graph-fraud}
\end{figure}

One step in this direction is to enable the detection of patterns of certain kinds, for example, cycles and cliques (or densely connected sub-graphs). As additional examples, detecting a graph pattern where a series of transactions eventually merge back to a single account might indicate a Ponzi scheme \cite{DBLP:journals/tcasII/JinJZWX22,DBLP:conf/coin-ws/FratricSEK22,ponzichallenges}, while a transaction graph with a ``pyramidoid" layered structure might indicate a pyramid scheme \cite{pyramidscheme}. Such patterns are illustrated in Figure \ref{fig:graph-fraud}.
These and other patterns are known to be of great interest as litmus tests of fraudulent activity. 
 

In the case of cross-border CBDC payments, payment intermediaries may also participate in the MPC protocol since they can keep a record of the sender and the receiver bank and the respective amounts for each transaction unless there is a privacy-preserving implementation where the intermediary does not learn any such information and only executes CBDC conversions/transfers. Also, when following the outsourcing approach discussed previously, the validator committees across multiple CBDC networks can work together in a hierarchical fashion to aggregate data in a scalable fashion, akin to sharded MPC protocols in  \cite{EPRINT:DKMSZ17}.

\subsection{Homomorphic Encryption}
\subsubsection{Background.}
Homomorphic encryption is a special method of encryption
that enables computation over encrypted data~\cite{EPRINT:ABCGJR15,sealcrypto}. HE can be partially homomorphic (i.e., additive or multiplicative, which only enable addition or multiplication of ciphertexts respectively), or fully homomorphic (which enables both mathematical operations).
In both cases, these resulting computations over encrypted data, when decrypted, should result in the same output as if the same mathematical operations have been performed over the respective plaintext data.

\subsubsection{HE in Financial Data Sharing.}
Homomorphic encryption can be utilized when multiple encrypted inputs from banks are outsourced to a single server in order to jointly perform some computation over the joint data. In this setting, threshold fully homomorphic encryption (TFHE)~\cite{EC:AJLTVW12}, a special type of fully homomorphic encryption, can be particularly useful, as it enables the banks to jointly create a common public key with the respective secret key shared among them, and then decrypt the computed homomorphic ciphertext without learning anything but the resulting plaintext.
This method can serve use cases such as computing macroeconomic statistics; for example, the banks can encrypt the value of their transaction totals using TFHE, outsource the computation of the encrypted average value to a server, and then decrypt the average value as the final result.

\subsection{Zero Knowledge Proofs}
\subsubsection{Background.} 
Zero-knowledge proofs are cryptographic methods that enable a party, having the role of the prover, 
to prove to another party, which has the role of the verifier, 
that a given statement is true, without revealing any additional information other than the statement is indeed true~\cite{STOC:BluFelMic88}.  For example, a proving party can compute $H(x)=y$ where $H$ is a hash function, and given $y$ as a publicly known value, would prove knowledge of the preimage $x$ but without revealing any information about $x$. 
Note that research over the last years has significantly advanced the practicality of ZKPs, for example, zk-SNARKs\footnote{Stands for zero-knowledge succinct non-interactive argument of knowledge}~\cite{SP:PHGR13,EC:Groth16,EPRINT:GabWilCio19} which have a complex mathematical basis, are proofs that are short in size and easy to verify.


\subsubsection{ZKPs in Financial Data Sharing.}
ZKPs are typically used in payment systems that provide end-to-end privacy (e.g., Zcash or Zether~\cite{SP:BCGGMT14,FC:BAZB20}). However, for the purposes of financial data sharing with cross-institutional privacy, such proofs can be utilized mostly for protection against  malicious adversaries. 
In particular, such proofs can be helpful towards addressing one of the major challenges for the computation of financial data using MPC, where malicious clients can bias their input, and thus the output, by submitting maliciously outlier input, i.e., data points that are created on purpose to deviate from the general trend of the overall data. If the data was in plaintext, such data points could be detected and excluded using outlier detection algorithms common in data analytics. However, encrypted data makes it challenging to detect such behavior. 
For instance, ZKPs are used in Verifiable Secret Sharing~\cite{FOCS:CGMA85}), while Prio \cite{DBLP:conf/nsdi/Corrigan-GibbsB17} offers a mechanism for data providers to prove to an aggregator that their data points are free from statistical biases without revealing their data in plaintext, proposes an efficient proof mechanism called secret-shared non-interactive proofs (SNIPs), and shows the feasibility of doing computation over private data with a throughput of thousands of submissions per second.
Note that similar approaches could potentially be employed in Machine Learning or Federated Learning solutions to ensure the integrity of shared financial data.

Finally, zk-SNARKs can also be utilized towards compressing large amounts of data containing financial transactions. For instance, again in the cryptocurrency space, Mina~\cite{EPRINT:BMRS20a} uses zk-SNARKs in a recursive fashion in order to compress the entire blockchain state in a few kBytes, while  zk-rollups~\cite{rollups} also use them as a layer-2 scalability solution to move data and computation off-chain. 

\subsection{Federated Learning}
\subsubsection{Background.} 
Federated learning is an approach for scaling and enhancing the data privacy of machine learning algorithms~\cite{DBLP:journals/corr/abs-1912-04977}. FL enables the decentralized execution of the training phase in machine learning to happen across several clients holding local data samples. 
This process is facilitated by a server, which however does not learn anything about the client's private data. For example,  Gboard~\cite{DBLP:journals/corr/abs-1811-03604} learns new words and phrases for each individual device depending on the typed text with the help of servers, but without sending plaintext information to the servers. Furthermore, FL settings are typically categorized into \emph{cross-device} and \emph{cross-silo} FL~\cite{DBLP:journals/corr/abs-1912-04977}, where cross-device considers a very large number of devices (e.g., the Gboard example above) where those devices can be periodically offline, while in the cross-silo setting, clients in a smaller scale (e.g., FI's) need to train a model based on their siloed data. The latter category is more applicable to privacy-preserving financial data sharing.


\subsubsection{FL for Financial Data Sharing.}
While solutions have been proposed to differentiate between benign and malicious actors using general-purpose machine-learning techniques (e.g. in cryptocurrencies \cite{DBLP:journals/tsmc/WuYLYCCZ22,DBLP:conf/cscml/KumarSHS20,DBLP:journals/ipm/FanFXC21,DBLP:conf/icufn/BaekOKL19,DBLP:journals/eswa/FarrugiaEA20}), these solutions do not come with privacy in mind, and the commercial banks of our example might not be willing to adopt them for this reason. Therefore, FL is a potential solution as a privacy-preserving machine-learning technique. In the context of payment systems, \cite{DBLP:journals/corr/abs-1909-12946} which combined graph-based machine learning with federated learning across multiple financial institutions, is among the handful of proposals. This work, however, still seems to be far from practical, as it does not consider sharing heterogeneous data (which is the typical scenario for commercial banks, especially for cross-border settings) and does not consider the risk of inference attacks (i.e., reverse engineering the learning model), which would expose private data. 
More recently, FL was considered in a competition hosted by the US and UK governments~\cite{petschallenge} as the core PETs to facilitate financial data sharing towards preventing financial crime. The competition setting considered the Society for Worldwide Interbank Financial Telecommunication (SWIFT) which provides the network 
for handling messages to execute international payments across different banks, as a natural centralized information hub, which can utilize account information from banks in a privacy-preserving way to detect ``anomalous" transactions.

In general, FL is a promising solution for the problem of privacy-preserving financial data sharing, even when assuming homogeneous data sets as a starting point (while still heterogeneous data sets can be considered \cite{https://doi.org/10.48550/arxiv.2110.10927}). Towards addressing inference attacks, the CBDC DLT validator committee members can potentially act as servers in the Federated learning setting, possibly by introducing an amount of noise as it aggregates the received models from the banks to defend against inference attacks at the cost of accuracy.

\section{The PET-as-a-Service Paradigm}


While PETs allow enterprises to improve their existing services by sharing data with each other, and offer new flows that are not traditionally possible due to the sensitivity of underlying data, deploying a PET protocol in practice often brings significant operational overhead to an enterprise. This mainly stems from the complexity of the new technology that requires hiring scientists and engineers familiar with the new technology as well as implementing, customizing, and auditing new software components according to business needs. 

Meanwhile, as the industry and government bodies are experimenting with the technology~\cite{Cryptogr25:online,GitHubfb:online,PrivacyP31:online,UKUSpriz68:online}, common PETs design patterns are emerging that can significantly reduce the cost of using the technology by making it available as ``off-the-shelf'' software and further as a service commercialized by technology providers. We refer to such a model as the \textit{PET-as-a-service (PETaaS)} paradigm. In this paradigm, a client can initiate the execution of a private data-sharing protocol via a simple API without getting exposed to the complexities of the underlying PET. In the financial scenario for example, an FI uses a PETaaS SDK to locally mask its input data, chooses the function to be computed, and submits the request to the PETasS provider along with standard parameters to control the execution such as its duration, the list of input providers and output receivers, etc. 

In turn, the PETaaS provider, after choosing the appropriate PET protocol based on the parameters chosen, selects a committee of decentralized nodes from a pool of node/input providers, initiates the setup needed for the respective PET (e.g., generating keys), and assigns a node to each FI by facilitating the transfer of the masked data from each FI to its assigned node. Then, the PETaaS provider orchestrates the needed computation within the node committee. After the computation is complete, the output is handled according to the specified policies. For example, the output might only be learned by the FI which initiated the computation, or it might be shared among a set of authorities. 

A PETaaS provider may either offer general computational resources to the underlying PET protocol execution (e.g., solely as a cloud node provider charging fees for the cycles/bandwidth offered) and/or provide business-specific input data with the network privately to earn sharing fees and/or access to the computation output. 





\bibliographystyle{splncs04}
\bibliography{mybibliography,abbrev3,crypto}

\section*{Disclaimer}
\emph{Case studies, comparisons, statistics, research, and recommendations are provided “AS IS” and intended for informational purposes only and should not be relied upon for operational, marketing, legal, technical, tax, financial, or other advice.  Visa Inc. neither makes any warranty or representation as to the completeness or accuracy of the information within this document nor assumes any liability or responsibility that may result from reliance on such information. The information contained herein is not intended as investment or legal advice, and readers are encouraged to seek the advice of a competent professional where such advice is required.}

\emph{These materials and best practice recommendations are provided for informational purposes only and should not be relied upon for marketing, legal, regulatory, or other advice. Recommended marketing materials should be independently evaluated in light of your specific business needs and any applicable laws and regulations. Visa is not responsible for your use of the marketing materials, best practice recommendations, or other information, including errors of any kind, contained in this document.}

\end{document}